\documentclass[aps,prb,twocolumn,showpacs,superscriptaddress,floatfix,nofootinbib]{revtex4-1}
\usepackage{amsfonts,hyperref}

\usepackage{graphicx}
\usepackage{amsmath}
\usepackage{amssymb}
\usepackage{bm}
\usepackage{gensymb}

\newcommand{\hc}{\hat c}

\begin{document}
\title{Variational optimization with infinite projected entangled-pair states}

\author{Philippe Corboz}
\affiliation{Institute for Theoretical Physics, University of Amsterdam,
Science Park 904 Postbus 94485, 1090 GL Amsterdam, The Netherlands}

\date{\today}

\begin{abstract}
We present a scheme to perform an iterative variational optimization with infinite projected entangled-pair states (iPEPS), a tensor network ansatz for a two-dimensional wave function in the thermodynamic limit,  to compute the ground state of a local Hamiltonian. The method is based on a systematic summation of Hamiltonian contributions using the corner transfer-matrix method. Benchmark results for challenging problems are presented, including the 2D Heisenberg model, the Shastry-Sutherland model, and the $t$-$J$ model, which show that the variational scheme yields considerably more accurate results than the previously best imaginary time evolution algorithm, with a similar computational cost and with a faster convergence towards the ground state. 
\end{abstract}

\pacs{02.70.-c, 75.10.Jm, 71.10.Fd, 03.67.-a	}


\maketitle

\section{Introduction}
Understanding the collective phenomena in strongly correlated quantum many-body systems is one of the grand challenges in modern physics. For one-dimensional problems tremendous progress has been made thanks to the well-known density matrix renormalization group (DMRG) method.~\cite{white1992} DMRG has an underlying variational ansatz, called matrix product states (MPS), in which the wave function is efficiently represented by a product of matrices.~\cite{Verstraete08,schollwoeck2011}  The accuracy of the ansatz can be systematically controlled by the bond dimension $D$, corresponding to the linear size of the matrices. For two-dimensional systems, however, DMRG suffers from an exponential scaling of the computational cost with the linear system size.~\cite{stoudenmire12}

Progress in quantum information theory, in particular a better understanding of entanglement in quantum many-body systems, has led to the generalization of MPS to higher dimensions - called projected entangled-pair states~\cite{verstraete2004,murg2007,Verstraete08} or tensor product states.~\cite{nishino01,nishio2004,daniska15} These states reproduce an area law scaling of the entanglement entropy which typical ground states of local Hamiltonians fulfill,\cite{eisert2010} and thus  provide an efficient variational ansatz overcoming the exponential scaling of 2D DMRG with system size. A particular powerful ansatz is infinite PEPS~\cite{jordan2008} (iPEPS) which represents a wave function directly in the thermodynamic limit, and thereby minimizes finite-size and boundary effects.~\footnote{Another class of efficient tensor network states for 1D and 2D systems is the multi-scale entanglement renormalization ansatz\cite{vidal2007-1,evenbly2009-2} (MERA)}

In recent years 2D tensor network methods have become more and more powerful tools for the study of 2D fermionic and frustrated systems, see e.g. Refs.~\onlinecite{corboz11-su4,wang11_j1j2, Zhao12,  Corboz12_su4, xie14, Corboz13_shastry, gu2013, Osorio14, corboz14_tJ,corboz14_shastry, picot15a, picot15,corboz16,nataf16} and references therein. For example, it has been demonstrated that iPEPS provides better variational energies than state-of-the-art variational methods for the  \mbox{$t$-$J$} and the 2D Hubbard model.~\cite{corboz14_tJ,corboz16} Thanks to largely unbiased simulations iPEPS also played the key role in understanding the magnetization process in SrCu$_2$(BO$_3$)$_2$~\cite{corboz14_shastry} which has been an intriguing open problem for many years. Thus, already today, 2D tensor networks are very powerful, however, it is important to realize that the algorithms can still be further improved, such that even higher accuracies can be reached.

The biggest challenge in a 2D tensor network algorithm is the \emph{optimization} of the tensors, i.e. finding the best variational parameters stored in the tensors to have the best approximation of the ground state of a given input Hamiltonian. For iPEPS this is commonly done by performing an imaginary time evolution (ITE),~\cite{jordan2008, jiang2008, corboz2010, phien15} where a random initial iPEPS gets projected onto the ground state. The ITE cannot be done in an exact way, but requires truncating the bond indices of the iPEPS at each time step. Besides the truncation error there is an additional error coming from the Trotter-Suzuki decomposition of the ITE operator. In order to reduce the error one has to use very small time steps $\tau$, which is computationally not efficient, because many  steps are required to reach convergence.

In this paper we present an alternative optimization method for iPEPS based on a variational energy minimization. Similarly as in DMRG the idea is to perform sweeps over all tensors in the ansatz, and at each step one minimizes the energy with respect to a tensor while keeping the other tensors fixed. We show that this optimization method does not only converge faster to the ground state, but, surprisingly, it also yields considerably more accurate results  than the  best ITE algorithm (the so-called full update), even when taking the limit $\tau \rightarrow 0$.

A related scheme has already been applied to finite PEPS previously,~\cite{verstraete2004, murg2007,Verstraete08} however, for iPEPS there are two additional complications: (1) in order to perform the energy minimization one needs to take into account an infinite number of Hamiltonian contributions. 
In this paper we show how a systematic summation of all these contributions can be achieved using the corner-transfer matrix (CTM) method.\cite{nishino1996,nishino97, orus2009-1} [Alternatively the summation can also be done based on "channel environments"~\cite{Vanderstraeten15} or by representing the Hamiltonian as a projected entangled-pair operator,~\cite{frowis10} which is closely related to variational tensor network schemes which have been developed for 3D classical systems.~\cite{nishino01,gendiar03}] (2)~Because in iPEPS each tensor appears an infinite number of times in the ansatz (instead of only once as in finite PEPS), the energy optimization for a given tensor is a highly non-linear problem. We present a practical scheme to deal with this issue, which  provides a good convergence to the ground state.

The paper is organized as follows: in the next section we provide a short introduction to the iPEPS ansatz, the CTM method, and the optimization based on ITE. In Sec.~\ref{sec:varopt} we first present the main idea of the iterative variational optimization, then explain how to use the CTM method for the systematic summation of Hamiltonian contributions, and finally discuss practical schemes for the optimization. In  Sec.~\ref{sec:benchmarks} we present benchmark results for the 2D Heisenberg model, the Shastry-Sutherland model, and the $t$-$J$ model, to demonstrate the performance of the variational approach compared to results based on ITE. Finally,  we discuss and summarize our findings in Sec.~\ref{sec:summary}. In addition, in appendix~A we explain how to implement a two-site variational optimization which can be used complementary to the one-site update discussed in the main text. 
 


\section{Introduction to iPEPS}
\subsection{iPEPS ansatz}

\begin{figure}
\includegraphics[width=0.9\columnwidth]{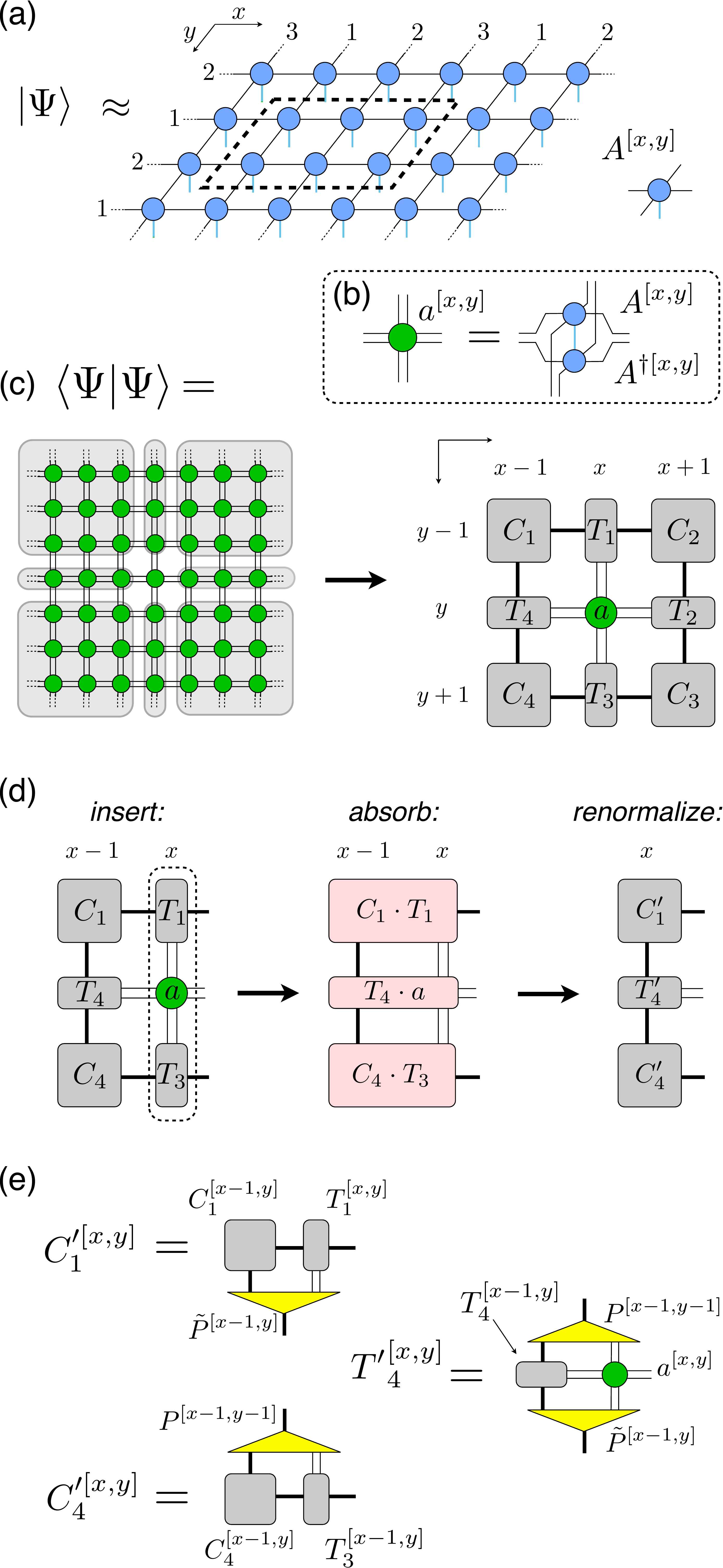}
\caption{(a) iPEPS ansatz with a $3\times 2$ unit cell of tensors which is periodically repeated in the lattice. (b) The reduced tensor $a^{[x,y]}$ is obtained from contracting an iPEPS tensor $A^{[x,y]}$ with its conjugate $A^{\dagger[x,y]}$ along the physical leg. (c)~The norm $\langle \Psi | \Psi \rangle$ is represented as an infinite square lattice network of reduced tensors. The CTM approach yields the \emph{environment} tensors surrounding a bulk tensor $a^{[x,y]}$ where the corner tensors $C_1$, $C_2$, $C_3$, $C_4$ take into account a quarter-infinite system, and the edge tensors $T_1$, $T_2$, $T_3$, $T_4$ an infinite half-row or half-column of the system. 
(d) A left move is done by inserting a new column of tensors, multiplying the tensors to the left, and performing a renormalization step (this is done for all coordinates $y$). (e)~Diagrams to compute the updated corner and edge tensors $C'_1$, $C'_4$, $T'_4$ at coordinate x (for all $y$ coordinates). Note that the coordinates are always taken modulo the unit cell size.
}
\label{fig:CTM}
\end{figure}

An iPEPS is an efficient variational tensor network ansatz for 2D ground states of local Hamiltonians in the thermodynamic limit~\cite{verstraete2004,jordan2008,nishino01,nishio2004, corboz2011} which obey an area law of the entanglement entropy.~\cite{eisert2010} It consists of a rectangular unit cell of tensors with one tensor per lattice site, $A^{[x,y]}$ where $[x,y]$ label the coordinates of a tensor relative to the unit cell of size $L_x \times L_y = N_T$, shown in Fig.~\ref{fig:CTM}(a). Each tensor has one physical index - carrying the local Hilbert space of a lattice site - and four auxiliary indices which connect to the nearest-neighbor tensors on a square lattice (more generally, an PEPS has $z$ auxiliary indices where $z$ is the coordination number of the lattice). The accuracy of the ansatz can be systematically controlled by the bond dimension $D$ of the auxiliary indices.

For a translational invariant state an ansatz with a single-tensor unit cell  can be chosen. However, if translational symmetry is spontaneously broken, a larger unit cell size  compatible with the periodicity of the ground state is required (for example, for an antiferromagnetic state two different tensors for the two sublattices are needed). Since the periodicity of the ground state is typically not known in advance one has to perform simulations with different unit cell sizes to determine which cell size leads to the lowest variational energy. Using different unit cells also offers the possibility to find different competing low-energy states (see e.g. Ref.~\onlinecite{corboz14_tJ}).

\subsection{Contraction of an iPEPS}
\label{sec:CTM}
In order to compute an expectation value of an observable $\hat O$ with respect to an iPEPS wave function $|\Psi\rangle$, the corresponding 2D tensor network representing $\langle \Psi | \hat O | \Psi \rangle$ has to be contracted in a controlled, approximate way. In this work we use a variant of of the corner-transfer matrix (CTM) renormalization group method,\cite{nishino1996, nishino97,orus2009-1} for arbitrary unit cell sizes~\cite{corboz2011,corboz14_tJ} which is summarized in the following.

Consider the problem of computing the norm of an iPEPS $\langle \Psi | \Psi \rangle$, which boils down to contracting the infinite 2D square lattice network of the reduced tensors $a^{[x,y]}$, shown in Fig.~\ref{fig:CTM}(c), where each $a^{[x,y]}$ is obtained from contracting $A^{[x,y]}$ with its conjugate tensor $A^{\dagger [x,y]}$, see Fig.~\ref{fig:CTM}(b). The goal of the CTM approach is to compute the four corner tensors $C_1$, $C_2$, $C_3$, $C_4$, and the four edge tensors $T_1$, $T_2$, $T_3$, $T_4$ for each coordinate $[x,y]$ in the unit cell, where each corner tensor represents a quadrant and the edge tensors a half-row (or half-column)       of the infinite 2D network. All these tensors together form the so-called environment, representing the infinite system surrounding a bulk site (or several bulk sites), as shown in Fig.~\ref{fig:CTM}(c). Once the environment has been computed, one can easily evaluate expectation values of local observables by introducing the corresponding operators in between the physical legs of the iPEPS tensors.

The environment tensors are computed iteratively by letting the system grow in all directions. One starts from an initial guess for the boundary tensors, either by initializing them randomly, or alternatively one can initialize them with the bulk tensors (by tracing out the auxiliary bonds on the edges). 
In the directional CTM approach~\cite{orus2009-1} one first performs a growth step on e.g. the left side of the system (called a left move), by introducing a new column of tensors,  multiplying them onto the left boundary tensors, followed by a renormalization step, see Fig.~\ref{fig:CTM}(d). 

In the renormalization step a bond dimension $\chi$ is kept at the boundary which controls the accuracy of the approximate contraction. There are different ways how to perform this renormalization step. Here we use a set of projectors $P$ and $\tilde P$,
 introduced in Refs.~\onlinecite{Wang11, Huang12} and first applied in the CTM method in Ref.~\onlinecite{corboz14_tJ}, to project from the enlarged space $\chi D^2$ down to a dimension $\chi$. These projectors are then used to compute the renormalized corner- and edge tensors, $C'_1$, $C'_4$, and $T'_4$, as shown in Fig.~\ref{fig:CTM}(e).

For a unit cell of size $L_x\times L_y$ one proceeds in the following way for a full left move (i.e. an absorption of the entire unit cell into the left boundary):
\begin{itemize}
\item Do for all $x \in [1,L_x]$
\begin{itemize}
\item Do for all $y \in [1,L_y]$
\begin{itemize}
\item Compute the projectors $P^{[x-1,y]}$ and $\tilde P^{[x-1,y]}$ (see Ref.~\onlinecite{corboz14_tJ} for details)
\end{itemize}
\item Do for all $y \in [1,L_y]$
\begin{itemize}
\item Compute the new renormalized corner tensors $C_1'^{[x,y]}$, $C_4'^{[x,y]}$, and edge tensor $T_4'^{[x,y]}$, as shown in Fig.~\ref{fig:CTM}(e)
\end{itemize}
\end{itemize}
\end{itemize}

After a full left move one proceeds with a full right-, top-, bottom-move in a similar way, and reiterates until convergence is reached (e.g. by checking the convergence of the energy with CTM iterations).

\subsection{Optimization based on imaginary time evolution}
\label{sec:ite}
In order to get an approximate representation of the ground state of a given Hamiltonian $\hat H$, the tensors need to be \emph{optimized}, i.e. one needs to find the best variational parameters stored in the tensors. 
In previous iPEPS simulations this has been done based on an imaginary time evolution (ITE) of an initial (e.g. random) state. Using  a Trotter-Suzuki decomposition the imaginary time evolution operator is split into a product of two-site operators, 
\begin{equation}
e^{-\beta \hat H} = e^{-\beta \sum_b \hat H_b} \approx \left( \prod_b \hat U_b \right)^n, \quad \hat U_b = e^{-\tau \hat H_b},
\end{equation}
where the product goes over all nearest-neighbor bonds~$b$ in the unit cell (assuming a Hamiltonian with only nearest-neighbor terms), $\hat H_b$ is the Hamiltonian term on bond $b$, and  $\tau=\beta/n$ is a small imaginary time step. The error of the Trotter-Suzuki decomposition decreases with the size of the time-step $\tau$.~\footnote{We usually use a second order decomposition in our simulations.} The ITE is then performed by sequentially multiplying the two-site operators $\hat U_b$ to the iPEPS and representing the resulting wave function again as an iPEPS with the same bond dimension, until convergence is reached. There exist different schemes to truncate of a bond. In the so-called simple update scheme the truncation is done based on a local singular value decomposition,~\cite{vidal2003-1,jiang2008,corboz2010} whereas in the full-update\cite{jordan2008,corboz2010} (or fast-full update~\cite{phien15}) the entire 2D wave function is taken into account for the truncation of a bond index. The simple update is computationally cheaper, but less accurate than the full update. 



\section{Variational optimization}
\label{sec:varopt}
\subsection{Basic idea}
\begin{figure}
\includegraphics[width=1\columnwidth]{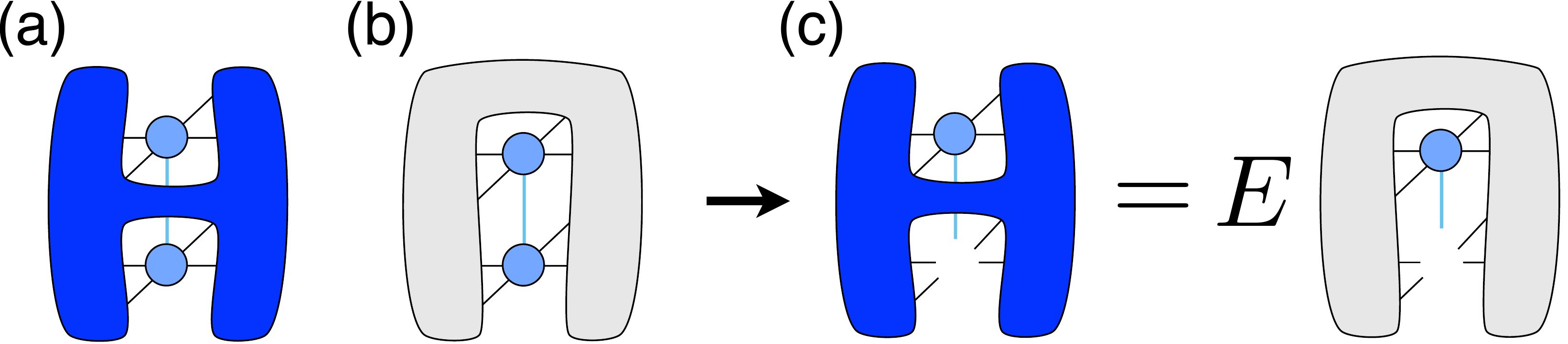}
\caption{(a) The H-shaped tensor is obtained by contracting the entire network representing $\langle \Psi | \hat H | \Psi \rangle$ except tensors $A$, and $A^\dagger$. (b) Similarly, the N-shaped tensor contains all tensors representing the norm $\langle \Psi | \Psi \rangle$ except tensors $A$, and $A^\dagger$, and can be obtained by contracting all environment tensors in Fig.~\ref{fig:CTM}(c) together. Minimizing the energy with respect to tensor $A$ boils down to solving the generalized eigenvalue problem shown in (c), by reshaping the H- and N-tensors into matrices, and $A$ into a vector.}
\label{fig:HN}
\end{figure}

Iterative variational optimization schemes are commonly used for MPS and finite PEPS,~\cite{white1992, schollwoeck2011,verstraete2004, murg2007,Verstraete08} but so far not for iPEPS (except for 3D classical systems~\cite{nishino01,gendiar03}). 
The main idea is to iteratively optimize one tensor after the other until convergence is reached. Optimizing a single tensor~$A$ (while keeping all other tensors fixed) boils down to minimizing the energy with respect to tensor A,
\begin{equation}
\label{eq:minE}
\min_A E(A) = \min_A \frac{\langle \Psi (A) | \hat H | \Psi (A) \rangle}{\langle \Psi (A) | \Psi (A) \rangle} =  \min_{\vec{A}} \frac{\vec{A}^\dagger {\bf H } \vec{A}}{\vec{A}^\dagger {\bf N} \vec{A}} 
\end{equation}
where the tensor $A$ and its conjugate have been reshaped into vectors. The matrices ${\bf N}$ and ${\bf H}$ correspond to the (reshaped) tensor network representing the norm and the expectation value of $\hat H$ excluding the tensor $A$ and its conjugate $A^\dagger$, respectively, see Fig.~\ref{fig:HN}. Minimizing with respect to $A^\dagger$ yields a generalized eigenvalue problem,
\begin{equation}
\label{eq:GEV}
\frac{\partial}{\partial \vec{A}^\dagger} \left(  \frac{\vec{A}^\dagger {\bf H} \vec{A}}{\vec{A}^\dagger {\bf N} \vec{A}}  \right) = 0, \quad \rightarrow \quad {\bf H} \vec{A} = E {\bf N} \vec{A}.
\end{equation}
The eigenvector $\vec{\tilde{A}}$ with lowest eigenvalue $\tilde E$ provides the solution to the local minimization problem, and the updated tensor $A'$ is obtained by reshaping $\vec {\tilde{A}}$ back to a tensor.

The main challenge of such a scheme for iPEPS is the computation of the matrix ${\bf H}$ which consists of an infinite sum of the expectation values of all Hamiltonian terms. In the following we explain how to obtain ${\bf H}$ using the CTM method, in a similar way as we computed the environment for the norm, ${\bf N}$, discussed in Sec.~\ref{sec:CTM}. The second complication comes from the fact that in iPEPS a tensor $A$ is not appearing only once in the ansatz (unlike in finite PEPS), but actually ${\bf H}$ and ${\bf N}$ also depend on $A$, making each step a highly-nonlinear optimization problem. We present a practical scheme dealing with this issue in Sec.~\ref{sec:practical} further below.


\subsection{Systematic summation of Hamiltonian terms with the CTM method}
\label{sec:ECTM}
\begin{figure}
\includegraphics[width=0.9\columnwidth]{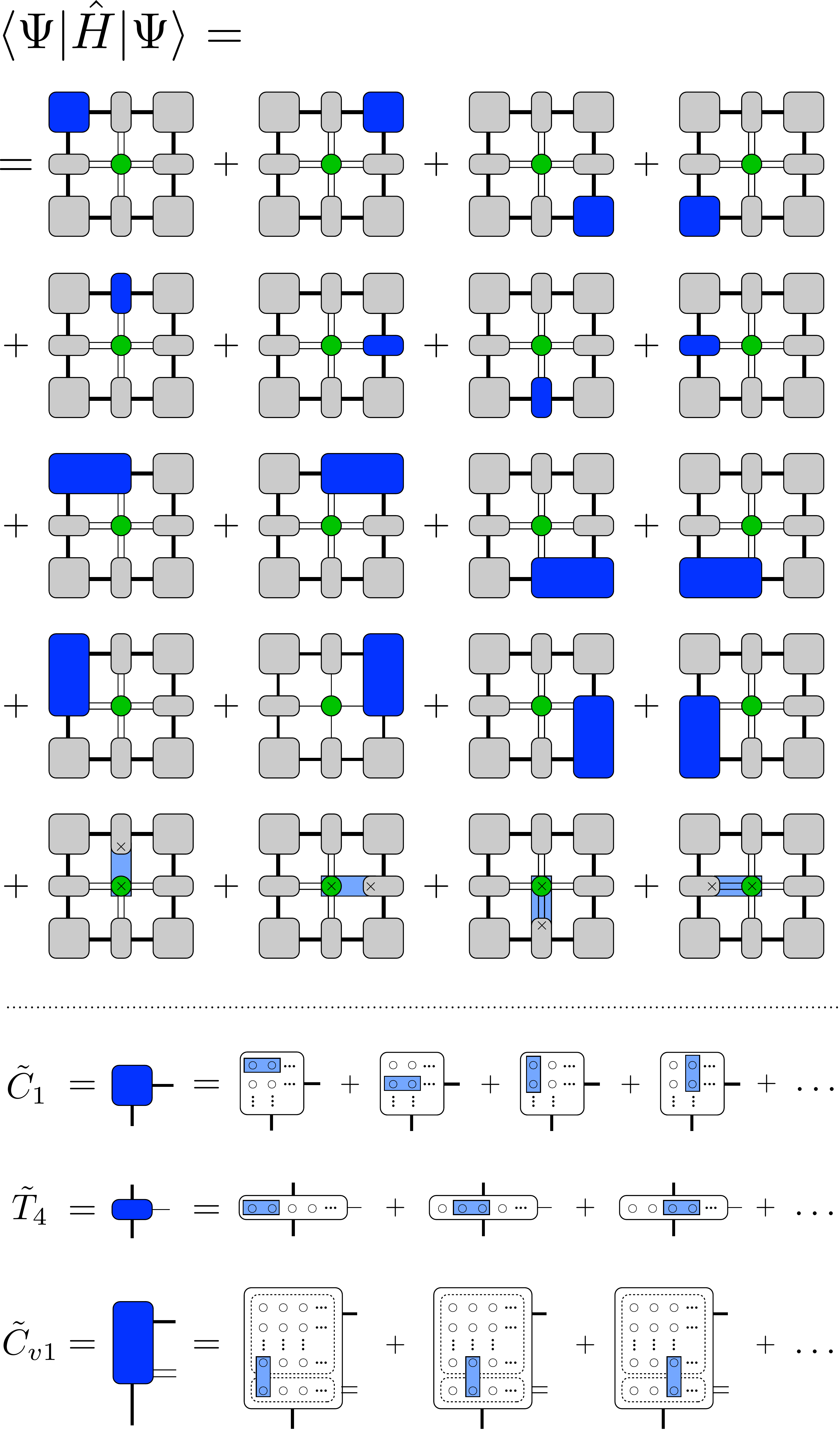}
\caption{(Color online) Representation of the expectation value of the Hamiltonian, where the blue tensors contain sums of local Hamiltonian terms, as illustrated in the bottom part of the figure. For example, the corner tensor $\tilde C_1$ contains all contributions of local Hamiltonian terms in the upper left corner of the infinite system, whereas the edge tensor $\tilde T_4$ contains all contributions from an infinite half-row, as depicted in the bottom part of the figure. The vertical corner tensor $\tilde{C}_{v1}$ takes into account all Hamiltonian terms located between the corner $C_1$ and the edge tensor $T_4$, see bottom image (a similar definition holds for the horizontal corner tensors $\tilde{C}_{h1}$). All the other dark-blue tensors on the other corners/edges are defined in a similar way. Finally, there are four remaining Hamiltonian terms (light blue) between the center site and its nearest neighbors. The 'x' on top of a tensor indicates that the Hamiltonian term is connected to the corresponding physical legs which are not shown in this top-view.}
\label{fig:Hsum}
\end{figure}

The CTM method discussed in \ref{sec:CTM} provides a convenient way to compute the norm (and local expectation values) by using the environment tensors, as shown in Fig.~\ref{fig:CTM}(c). The expectation value $\langle \Psi | \hat H | \Psi \rangle$, which is an infinite sum,  can be computed in a similar way by introducing new type of environment tensors which we call H-environment tensors, shown in dark blue in Fig.~\ref{fig:Hsum}. 

Each H-environment tensor consist of a sum of Hamiltonian contributions. For example, the corner tensor $\tilde C_1$ contains all contributions from Hamiltonian terms acting on the infinite upper-left part of the system (see lower panel in Fig.~\ref{fig:Hsum}). Similarly, $\tilde T_4$ contains all Hamiltonian terms acting on the corresponding infinite half-row. We further introduce horizontal and vertical corner tensors, denoted by $\tilde C_{h1}$ and $\tilde C_{v1}$, respectively, for the upper left corner. These tensors take into account Hamiltonian terms which connect sites located in the corner $C_1$ and edges $T_1$ or $T_4$, respectively (see bottom of Fig.~\ref{fig:Hsum}). Similar tensors are also defined for the other corners. Finally, we also have to sum up the local Hamiltonian terms connecting the center site with its four nearest neighbors (located on the four edge tensors). With this, the sum represented in Fig.~\ref{fig:Hsum} takes into account all Hamiltonian terms.

\begin{figure}
\includegraphics[width=0.9\columnwidth]{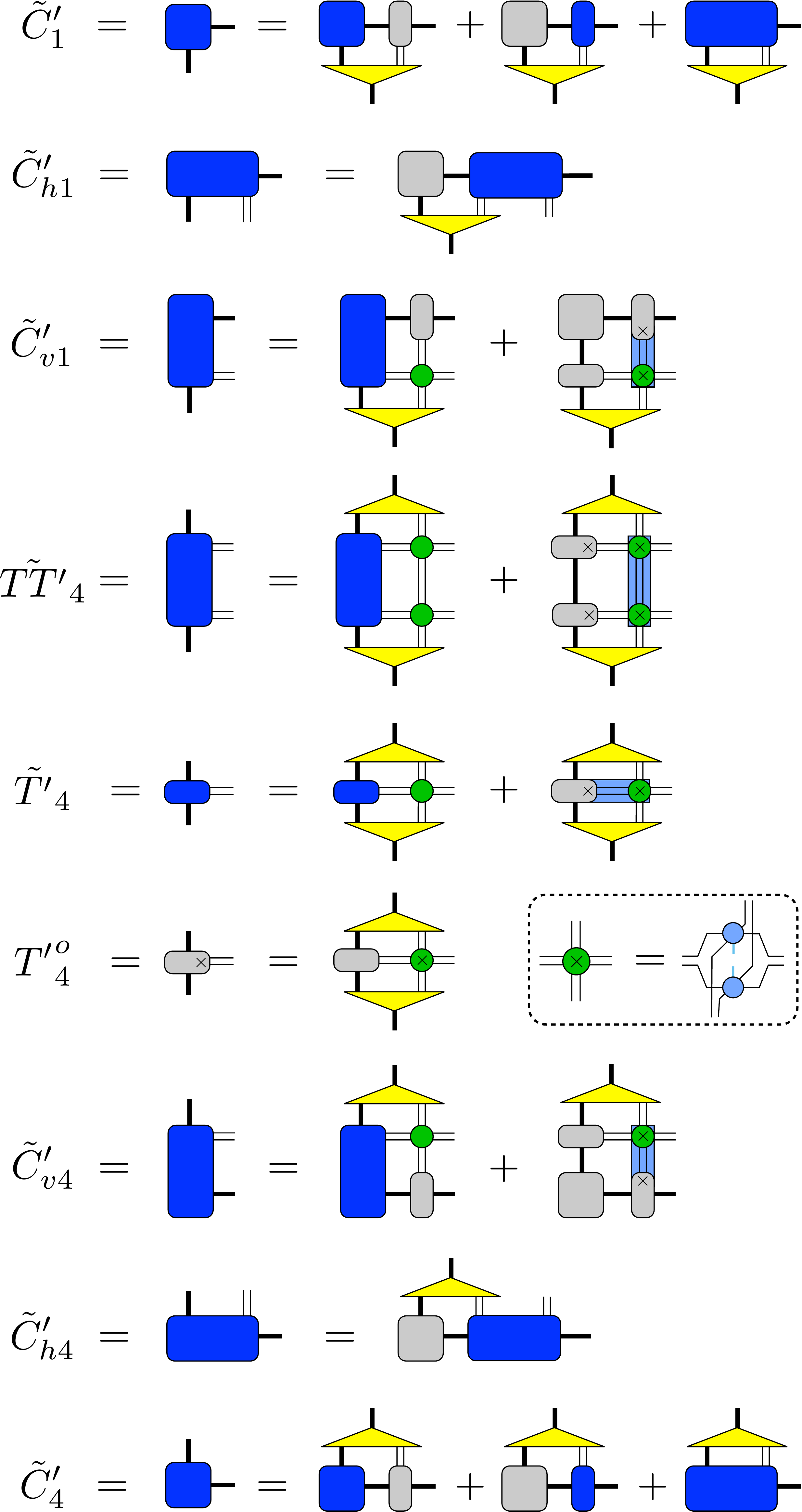}
\caption{All relevant diagrams to perform a left-move to update the H-environment tensors in the CTM method. Coordinates of the tensors relative to the unit cell have been omitted for simplicity. The projectors to perform the renormalization step (yellow triangles) are the same as the ones computed for the norm. The tensor $\tilde{TT}_4$ contains Hamiltonian terms connecting the sites between two edges $T_4$. The tensor $T_4^o$ is an edge where the physical legs of the right-outermost bulk tensors are left open (such that a Hamiltonian term can be connected to it). Similar diagrams are defined for a right-move, top-move and bottom-move. In this way one keeps track of all nearest-neighbor Hamiltonian terms in a systematic way.}
\label{fig:eleft}
\end{figure}

The H-environment tensors can be computed in a systematic way within the regular CTM method, as shown in Fig.~\ref{fig:eleft} for a left-move. Importantly, the H-environment tensors are renormalized in the same way as the norm-environment tensors, i.e. using the same projectors $P$ and $\tilde P$. In this way the indices of the H-environment tensors match with the ones from the norm-environment tensors, and thus different diagrams as shown in Fig.~\ref{fig:eleft} can simply be added.\footnote{If one would compute a separate set of projectors to renormalize the bond indices adjacent to an H-environment tensor, one would also loose translational invariance in the environment, since e.g. the $T$-environment tensors adjacent to a $\tilde T$ tensor would also get altered, and the number of different tensors would diverge with increasing CTM iterations which would make the scheme inefficient.}

Note that the $\tilde{TT}$ tensors, which include Hamiltonian contributions between two edge tensors, do not appear in the expectation value of the Hamiltonian shown in Fig.~\ref{fig:Hsum}. However, it is crucial to keep track of these tensors, since they add contributions to the $\tilde{C_{h}}$ and $\tilde{C_{v}}$ tensors, as for example shown in the second row in Fig.~\ref{fig:Hsum}, where the contributions in the $\tilde{TT_1}$ tensor are added to the  $\tilde{C'_{h1}}$ tensor.

We end this section with a three additional remarks: (1)~It is convenient  to store also the edge tensors where the physical legs of the outermost site are kept open, e.g. $T'^o_4$ shown in Fig.~\ref{fig:eleft}. These tensors can then be used to compute the local Hamiltonian terms (connecting to the center site) shown in Fig.~\ref{fig:Hsum}.
(2)~The computation of the $\tilde{TT}$ edge terms has a relatively large computational cost of $O(\chi^3 D^6)$ compared to the other terms. This is the same complexity~\footnote{By introducing further approximations the projectors can also be computed with a computational cost of $O(\chi^3 D^4)$} as for the computation of the projectors $P$ and $\tilde P$.~\cite{corboz14_tJ}  One way to reduce the complexity of the $\tilde{TT}$ term is to split it in the middle into two parts using an SVD, and keeping only a bond dimension of $O(\chi)$ between the two parts. (3) In some implementations of the CTM algorithm one normalizes the environment tensors in a certain way after each step (e.g. division by the largest element of a tensor) in order to keep the numbers in the tensors bounded. In this case one has to make sure for consistency that the same normalization is used also for the H-environment tensors (i.e. the same normalization factor has to be used e.g. for $C_1$ and $\tilde C_1$).


\subsection{Practical schemes}
\label{sec:practical}
With the CTM approach discussed in the previous sections we can compute the $\bf{H}$ and the $\bf{N}$ matrices and solve the generalized eigenvalue problem \eqref{eq:GEV} for the eigenstate $\tilde A^{[x,y]}$ with lowest energy eigenvalue. In finite PEPS, where each tensor appears only once, this provides the best solution at the current iteration. In iPEPS, however, each tensor $A^{[x,y]}$ appears infinitely many times, and thus replacing each tensor $A^{[x,y]}$ by the solution $\tilde A^{[x,y]}$ might not be the optimal choice. This is because both $\bf{H}$ and $\bf{N}$ also depend on $A^{[x,y]}$, making Eq.~\eqref{eq:minE} a highly-nonlinear problem (instead of a quadratic one). 

One could solve the minimization problem \eqref{eq:minE}, e.g., by a conjugate-gradient method. Here we use a different strategy, which turns out to work well in practice: we solve the generalized eigenvalue problem, but instead of using the solution $\tilde A^{[x,y]}$ we take a linear combination  with the previous tensor $A^{[x,y]}$,
\begin{equation}
A'(\lambda)^{[x,y]} =  \tilde A^{[x,y]} \sin \lambda \pi -  A^{[x,y]} \cos \lambda \pi.
\end{equation}
We then optimize the energy $E(\lambda)$ with respect to the single parameter $\lambda \in [0.5,1.5]$,~\footnote{A similar mixing was also used in Refs.~\onlinecite{nishino01,gendiar03} in the variational optimization schemes for 3D classical systems, however, without performing a second minimization step with respect to the mixing parameter.} which in principle can be done by standard minimization solvers. For each evaluation of $E(\lambda)$ one has to recompute the environment for the norm (typically a few iterations starting from the previous environment is accurate enough), and evaluate all local Hamiltonian terms. For this reason it is desirable to keep the number of function evaluations of $E(\lambda)$ low. We made good experience with the following scheme:
\begin{itemize}
\item 
Compute $E(1)$ (corresponding to the previous energy with the old tensor $A' = A$), and $E(0.5)$ (corresponding to the energy with $A' = \tilde A$).
\item 
If $E(0.5)<E(1)$, take $A' = \tilde A$ as solution and exit 
\item
Define an initial step size $\Delta_0$ (e.g. $\Delta_0=0.1$), and a tiny step size $h$ (e.g. $h=10^{-4}$)
\item
If $E(1+h)<E(1)$, set $\Delta = \Delta_0$, else $\Delta = - \Delta_0$
\item
For $iter = 1$ to $maxiter$
\begin{itemize}
\item 
If $E(1+\Delta) < E(1)$ accept solution~\footnote{Some (large) values of $\Delta$ may result in a ''pathological'' state where the CTM scheme does not properly converge  anymore and where the norm of the state becomes very small. Such solutions should not be accepted since they may lead to stability problems (see also discussion in Ref.~\onlinecite{gendiar03}).} with \mbox{$\lambda = 1+\Delta$} and exit
\item 
else $\Delta = \Delta / 2$
\end{itemize}
\end{itemize}

With this scheme typically only a few evaluations of the energy are required. The algorithm stops as soon as a lower-energy solution is found. This does not provide the optimal $\lambda$ at each iteration, but in practice this does not seem to matter since in the end we are interested in the global minimum after many sweeps, and not the "local" optimum at each iteration.

Finally, we repeat the minimization for each tensor in the unit cell, i.e. for all coordinates $[x,y]$, and reiterate until the desired convergence in the energy is reached.

For both computations of the $H$-environment and the norm-environment we can start from the  environment from the previous iteration (similarly as in the fast-full update~\cite{phien15}), so that only a few additional CTM iterations are needed at each step.\footnote{After many iterations it can be useful to reinitialize the environments in some cases, and to recompute the environments from scratch.}


\section{Benchmark results}
\label{sec:benchmarks}
In this section we present a series of  benchmark results, ranging from a standard problem (the Heisenberg model) to challenging cases, including the Shastry-Sutherland- and the $t$-$J$ model. In all  cases we show that the iPEPS results for each bond dimension can be considerably improved with the variational optimization,  the energy as well as  order parameters. For the larger $D$ simulations we have exploited the $U(1)$ symmetry of the models in order to increase the efficiency of the calculation.~\cite{singh2010,bauer2011}


\subsection{Heisenberg model}
As a first example we consider the two-dimensional S=1/2 Heisenberg model on a square lattice with Hamiltonian,
\begin{equation}
\hat H=J \sum_{\langle i,j \rangle}  \bm S_{i}\cdot \bm S_{j},
\end{equation}
where the sum goes over nearest-neighbor sites and $\bf S_i$ is a spin-1/2 operator on site $i$. Reference values are taken from Ref.~\onlinecite{sandvik2010} which are based on state-of-the-art Quantum Monte Carlo calculations. 
We use an iPEPS ansatz with 2 different tensors (one for each sublattice) to represent the ground state with antiferromagnetic order.

\begin{figure}
\includegraphics[width=1\columnwidth]{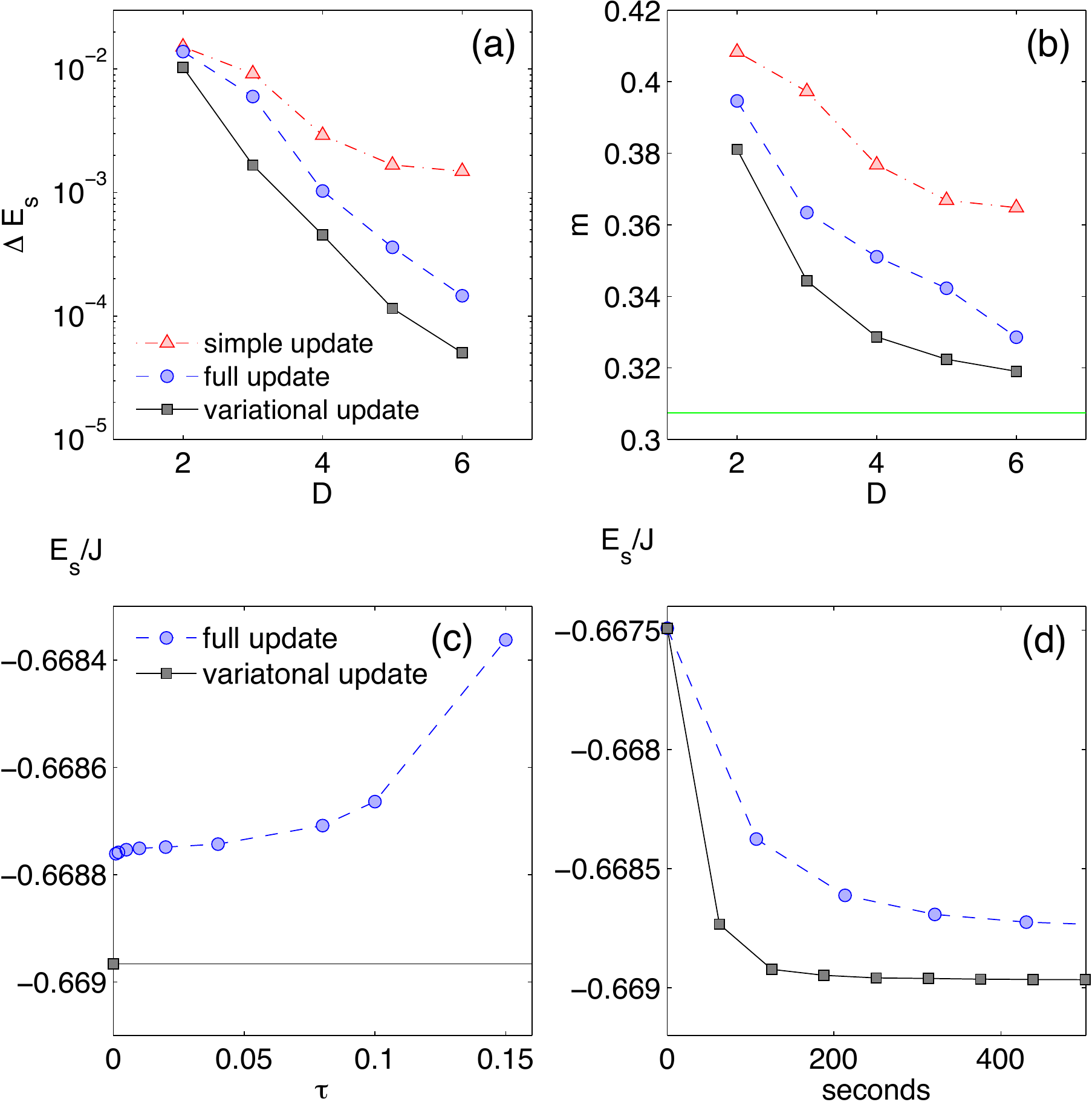}
\caption{(Color online) (a) Relative error of the energy per site of the 2D S=1/2 Heisenberg model as a function of the bond dimension $D$ obtained with three different optimization methods for iPEPS. (b) The order parameter (staggered magnetization) as a function of $D$, compared to the extrapolated QMC result in the thermodynamic limit (horizontal line). (c) Full update result as a function of the Trotter imaginary time step $\tau$ ($D=4$) compared to the variational result. (d) Evolution of the energy as a function of simulation runtime on a Macbook Pro laptop, for $D=4$ and $\chi=50$. The results from the variational optimization is shown at each sweep (squares), whereas the energy obtained with the full update is computed after every 30 time steps with $\tau=0.02$.
 }
\label{fig:heis}
\end{figure}

Figure \ref{fig:heis}(a) shows a comparison of the relative error of the energy as a function of the bond dimension $D$, obtained with the three optimization methods. One can clearly see that there is a substantial improvement when going from the simple update to the full update calculations as previously found. But interestingly, the results obtained with the variational optimization are even better, by roughly a factor 2! Also the order parameter, the staggered magnetization $m$ shown in Fig.~\ref{fig:heis}(b), is considerably improved.

It is important to point out that the difference between full- and the variational update is not due to the Trotter error in the imaginary time evolution evolution. In Fig.~\ref{fig:heis}(c) the dependence of the result for $D=4$ as a function of the Trotter step $\tau$ shows that even in the limit $\tau \rightarrow 0$ the full update does not yield the same accuracy as the variational approach. 
We believe that this has mainly to do with the fact that at every step a two site operator $\hat U_b$ (cf. Sec~\ref{sec:ite}) is  applied only to a \emph{single} bond in the middle of the system, which is then truncated in an optimal way resulting in updated tensors $A'$, $B'$. These tensors provide a locally optimal solution for that single bond, however, it is not guaranteed that they also provide a globally optimal solution when replacing the tensors everywhere in the ansatz. A globally optimal solution could be found by applying $\hat U_b$ to all bonds in the ansatz and then truncating all bonds simultaneously. However, such a scheme would be computationally much more expensive, because it would require to contract the time-evolved iPEPS with enlarged bond dimension, and thus it would not be very useful in practice.

The variational optimization is not only more accurate, but typically also converges faster to the lowest energy state. This is illustrated in Figure~\ref{fig:heis}(d) where we compare the performance of the full update with the variational update on a Macbook Pro laptop, for $D=4$, $\chi=50$. Starting from the simple update result already after one step with the variational scheme (taking 1 minute) one has a better result than with the full update in the long run-time limit. 

As a side remark on the performance we note that it is best to start with an initial state which is already close to the ground state, particularly for the large-$D$ simulations. For example, we can take the solution from the simple update (which is computationally very cheap) as an initial state for the full- or variational update scheme,\footnote{There are cases where simple update fails to provide the correct state, e.g. spin-liquid state,\cite{Osorio14} and does not provide a good initial state for these cases. It is thus important to crosscheck results also starting from random initial states or from full-update solutions.} or use the full-update solution (e.g. obtained with a large~$\tau$) as a starting point for a variational optimization. 
One can also use a converged solution with bond dimension $D-1$ as an initial state for a simulation with bond dimension $D$. In this case one can first perform a few full update steps (to increase the bond dimension from $D-1$ to $D$) and then continue with the variational update. Alternatively, the bond dimension can also be dynamically increased by using a 2-site update, see appendix \ref{sec:2site}.

Finally, it is interesting to compare our best result with 2D DMRG on cylinders from Ref.~\onlinecite{stoudenmire12}. Our lowest variational energy per site for $D=6$ ($\chi=250$)  is $-0.669408J$ which is very close to the exact Monte Carlo result, $-0.6694421(4)J$,\cite{sandvik2010} with a relative error of only 5e-5. This precision is comparable to a 2D DMRG calculation on a width-10 cylinder with $m=3000$ states. However, due to the exponential scaling of the required bond dimension with the cylinder width in DMRG, already a a width-12 cylinder with the same number of states is an order of magnitude worse than the iPEPS result. Furthermore, the iPEPS accuracy is obtained for the infinite system (infinite cylinder width), and we would actually expect an even higher accuracy with iPEPS on a finite cylinder.    
It is also remarkable to compare the number of variational parameters: A $D=6$ iPEPS has roughly 2.6e3 variational parameters per tensor (with 2 different tensors in the entire ansatz), whereas a $m=3000$ state has of the order of 1.8e7 parameters, i.e. a difference of four orders of magnitude per tensor. \footnote{In addition, a finite-size DMRG algorithm uses $L_x \times L_y$ different tensors, e.g. with $L_x =20$ and $L_y=10$ in Ref.~\onlinecite{stoudenmire12}, which increases the number of variational parameters by additional two orders of magnitude). We note that the effective number of parameters is reduced in both cases by exploiting the global U(1) symmetry in the model.} 
This illustrates that (i)PEPS offers a much more efficient representation of a 2D wave function than MPS, even for modest cylinder widths. 

In summary, the results from this section demonstrate that (1)~the full update actually \emph{fails} to reproduce the most optimal result for a given bond dimension $D$, even in the limit $\tau \rightarrow 0$, and (2)~that more accurate results can be obtained with the variational optimization, with typically a faster convergence towards the ground state.


\subsection{Shastry-Sutherland model}
We next move to a more challenging case: the Shastry-Sutherland model,~\cite{Shastry81} which is a frustrated spin system given by the Hamiltonian
\begin{equation}
H=J\sum_{\langle i,j \rangle}\bm S_{i}\cdot \bm S_{j}+J'\sum_{\langle \langle i,j \rangle\rangle}\bm S_{i}\cdot \bm S_{j}-h\sum_{i}S_i^z
\end{equation}
where the $\langle i,j \rangle$ bonds with coupling strength $J$ build an array of orthogonal dimers while
the  bonds with coupling $J'$ denote inter-dimer couplings, and $h$ the strength of an external magnetic field. This model is realized in the material SrCu$_2$(BO$_3$)$_2$ in which experiments have found a series of different magnetization plateaus.~\cite{Kageyama99,Onizuka00,kageyama00,Kodama02,takigawa04,levy08,Sebastian08,Jaime12,takigawa13,matsuda13} Recently, iPEPS has played the key role in understanding the structure of the states realized in these  plateaus.~\cite{corboz14_shastry}

At zero field, $h=0$, there exist a non-magnetic plaquette phase~\cite{Koga00, Takushima01, Chung01,Laeuchli02} in a narrow parameter range in between a dimer phase for $J'/J < 0.675(1)$ and an antiferromagnetic phase for $J'/J > 0.765(1)$.~\cite{Corboz13_shastry} Previous iPEPS simulations confirmed the existence of the plaquette phase,\cite{Corboz13_shastry} however, it was observed that due to the close adjacency of the other two phases it is difficult to converge into the plaquette phase with the full update (depending on the simulation setup and initial state used). For example, when starting from an antiferromagnetic state (using a simulation setup with one tensor per dimer), the full update fails to reproduce the plaquette state, and remains stuck in an antiferromagnetic state.

\begin{figure}
\includegraphics[width=1\columnwidth]{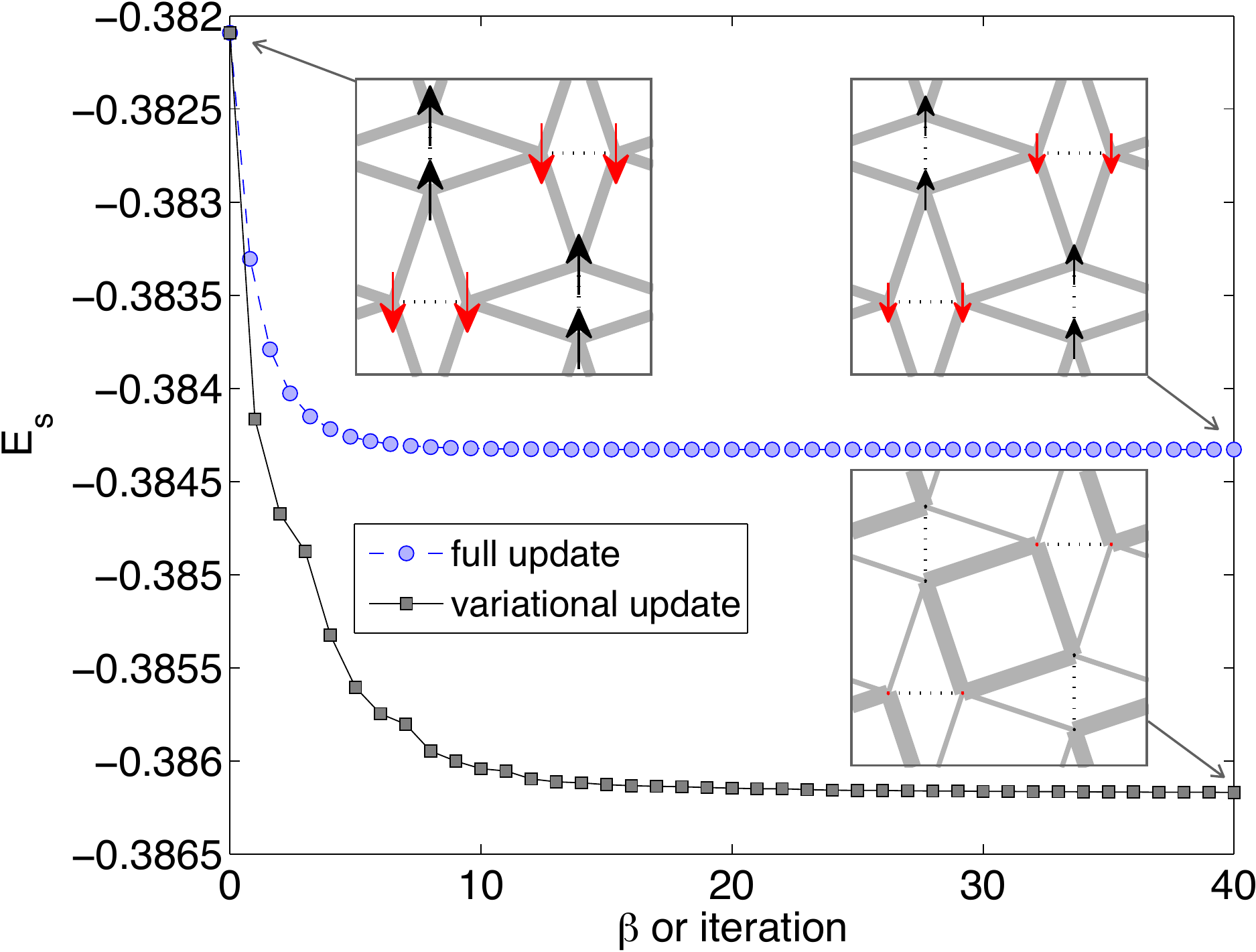}
\caption{Evolution of the energy as a function of imaginary time (full update) or iteration step (variational update), for $J'/J=0.7$ (in the plaquette phase) for $D=4$. Both simulations have been initialized  from an initial antiferromagnetic state which was obtained using the simple update (for $J'/J=0.78$). The initial state is shown in the upper left inset, where the length of the arrows are proportional to the magnitude of the local spin, and the thickness of the lines on the $J'$-bonds scales with the magnitude of the energy on that bond (i.e. the thicker a line the lower the energy). The variational update (squares) correctly reproduces a plaquette state, shown by the lower right inset, where the bond-energies around a plaquette are lower than on the other bonds, and the values of the local spins vanish. The full update (circles) fails to reproduce the state and yields an antiferromagnetic state with higher energy and slightly suppressed antiferromagnetic order compared to the initial state (upper right inset).}
\label{fig:plaquette}
\end{figure}

Here we use this challenging case as a benchmark for the variational optimization. Encouragingly, we find that even 
if the simulation is initialized in the wrong state, the variational optimization converges to the correct plaquette state. A comparison between the full update and variational results is presented in Fig.~\ref{fig:plaquette}, for $D=4$ and $J'/J=0.7$. Both simulations have been initialized with the same antiferromagnetic state obtained with the simple update (for $J'/J=0.78$). The resulting energy per site with the variational optimization, $E_s=-0.3862J$, is considerably lower than the full update result, $E_s=-0.3843J$. 

\begin{figure}
\includegraphics[width=1\columnwidth]{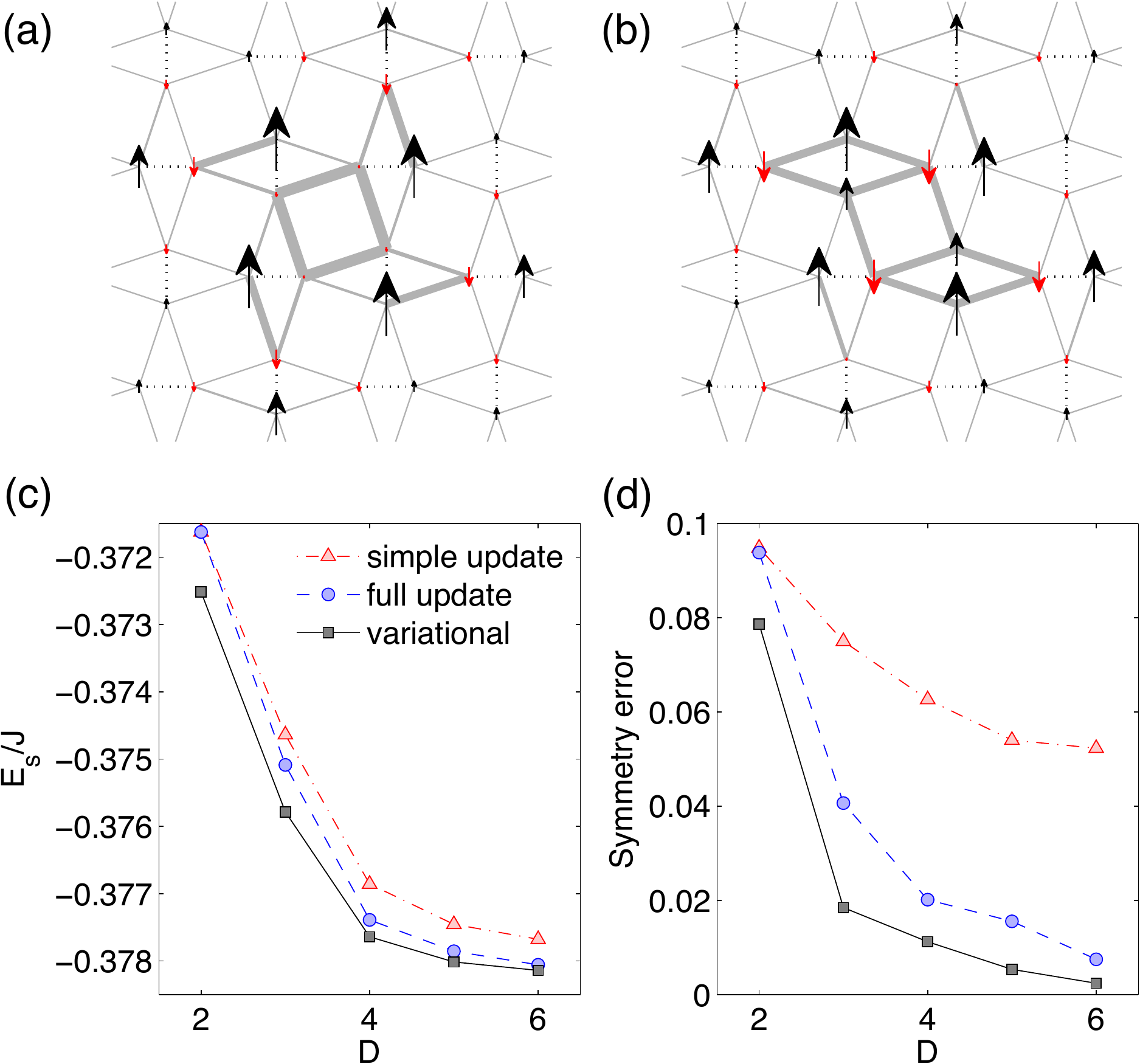}
\caption{(Color online) Results for the Shastry-Sutherland model in a finite magnetic field at magnetization $m=1/8$, obtained in a $4\times4$ unit cell of dimers. (a) Spin structure of a bound state of two triplet excitations exhibiting a $90\degree$ rotational symmetry. (b)~The rotational symmetry of a bound state gets strongly broken when using an inaccurate optimization such as the simple update (here for $D=5$). (c)~Comparison of the variational energies as a function of $D$ obtained with the three optimization methods. (d) Comparison of the symmetry error, quantifying the degree of the broken rotational symmetry.}
\label{fig:ssm_symmetry}
\end{figure}

In order to test the variational optimization method for larger unit cells we consider the Shastry-Sutherland model in an external magnetic field, i.e. with $h>0$. It has previously been  found with iPEPS that the magnetization plateaus correspond to crystals of bound states of triplet excitations,\cite{corboz14_shastry} which have a typical spin structure reminiscent of a pinwheel, shown in Fig.~\ref{fig:ssm_symmetry}(a) for a 4x4 unit cell of dimers (i.e. 16 different tensors). For details on the physics we refer to Ref.~\onlinecite{corboz14_shastry} and references therein. 
The bound state is symmetric under rotations by 90 degrees, however, an inaccurate optimization scheme like the simple update  fails to create a nicely symmetric state, as shown for example in Fig.~\ref{fig:ssm_symmetry}(b). The full update creates a more symmetric state, however, here we show that the variational optimization even provides a substantially better result.

To quantify the rotational symmetry breaking we define a "symmetry error" in the following way: On each site we determine the expectation value of the spin, and compute the standard deviation over all sites which are equivalent by symmetry. The "symmetry error" is then defined as the mean value of all the standard deviations. Thus, in a perfectly rotational invariant state, this error vanishes. Figure~\ref{fig:ssm_symmetry}(d) shows the results for the symmetry error as a function of $D$ obtained with the three optimization schemes. Again we find that the error is considerably lowered with the variational optimization compared to the simple- and full update. Consistently, the corresponding variational energies are also lower, as shown in Fig.~\ref{fig:ssm_symmetry}(c).

The results further demonstrate that with the variational optimization not only a better variational energy is obtained, but that the states (order parameters) are more accurately reproduced, even if they exhibit extended features requiring large unit cells.


\subsection{$t$-$J$ model}
Finally, we test the variational optimization also for a challenging fermionic system which requires a fermionic iPEPS ansatz. The formalism to apply 2D tensor networks to fermionic systems has been developed in several works,~\cite{Corboz10_fmera, kraus2010, pineda2010,  barthel2009, shi2009, Corboz09_fmera,corboz2010,pizorn2010,gu2010} which all describe the same fermionic tensor network ansatz. Here we follow the formalism from Ref.~\onlinecite{corboz2010} to take into account fermionic exchange statistics.

As a challenging example we consider the $t$-$J$ model which is an effective model of the Hubbard model in the strongly interacting limit, given by the Hamiltonian
\begin{equation}
\hat H= - t \sum_{\langle i,j \rangle \sigma} \left( \tilde{c}_{i \sigma}^{\dagger}\tilde{c}_{j\sigma}  + H.c.\right) +  J\sum_{\langle i,j \rangle}  \left( \hat S_i \hat S_j - \frac{1}{4} \hat n_i \hat n_j\right)  
\end{equation}
with $\sigma=\{\uparrow,\downarrow\}$ the spin index, $\hat n_i=\sum_\sigma \hc^\dagger_{i \sigma} \hc_{i \sigma}$ the electron density and $\hat S_i$ the spin $1/2$ operator on site $i$, and $\tilde{c}_{i\sigma}=\hc_{i\sigma} ( 1 - \hc^\dagger_{i \bar \sigma} \hc_{i \bar \sigma})$ hopping operators forbidding doubly occupied sites.

\begin{figure}
\includegraphics[width=0.9\columnwidth]{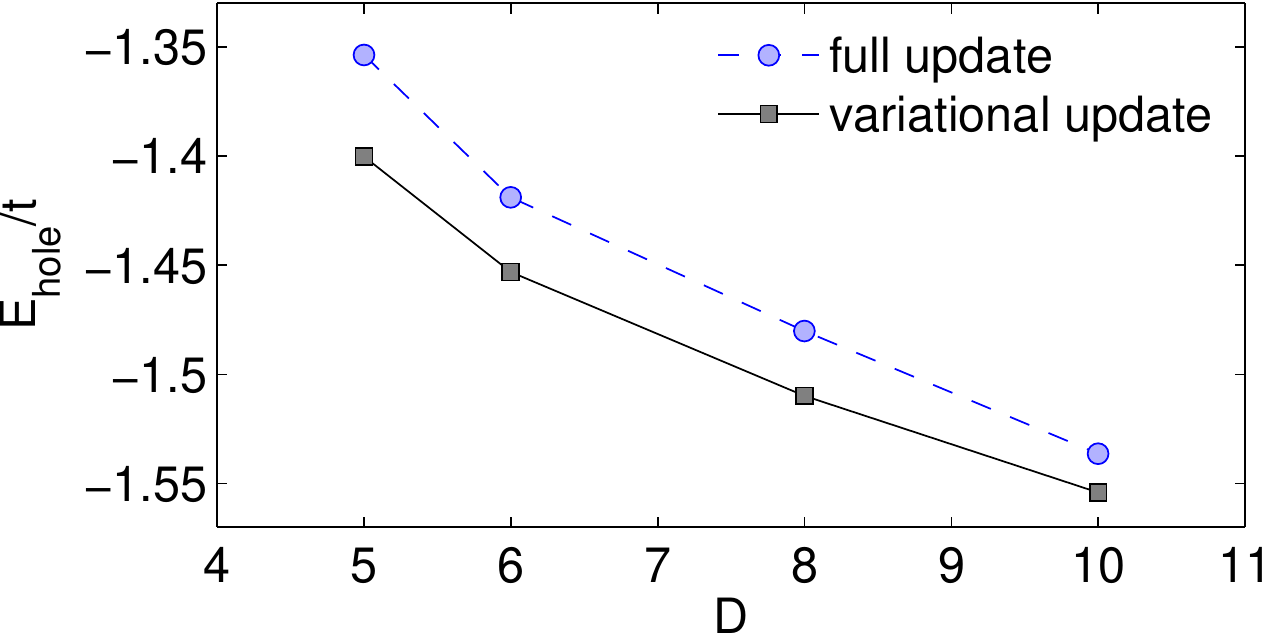}
\caption{(Color online) Energy per hole in the $t$-$J$ model for $J/t=0.4$ and doping $\delta=0.12$. For each value of $D$ the energies are substantially improved by the variational optimization.}
\label{fig:tJ}
\end{figure}

The $t$-$J$ model has been previously studied with iPEPS using the simple\cite{corboz2011} and full update.\cite{corboz14_tJ} Here we show that we can substantially improve the variational energies  by just adding a few (3-5) variational update steps on top of the full-update results from Ref.~\onlinecite{corboz14_tJ}. The results are presented in Fig.~\ref{fig:tJ} for $J/t=0.4$ and hole density \mbox{$\delta=0.12$}.

Thus, this example further illustrates the usefulness of the variational optimization to increase the accuracy of iPEPS simulations for challenging problems. Similar improvements can also be obtained for the 2D Hubbard model (not shown).~\cite{corboz16} 

\section{Summary and outlook}
\label{sec:summary}
We have presented a variational optimization scheme for iPEPS, based on a systematic summation of Hamiltonian terms using the corner transfer matrix method. For a given bond dimension $D$ the scheme yields a higher accuracy for the energy and order parameters than the previously best results based on an imaginary time evolution using the full update. For example, for the Heisenberg model with bond dimension $D=6$ the relative error of the energy is only 5e-5 (without using any extrapolation) compared to the extrapolated Quantum Monte Carlo result. The variational approach not only yields a higher accuracy but it can also help to speed up the convergence towards the ground state. We have also presented a challenging case where the full update failed to reproduce  the correct ground state in the Shastry-Sutherland model, whereas the variational optimization converged to the right plaquette state. 

We note that a similar approach can also be used for Hamiltonians with next-nearest neighbor terms with only minor modifications. In principle, also longer ranged Hamiltonians can be simulated which requires keeping track of additional H-environment tensors.

The results for the $t$-$J$ model provide further evidence that with the variational scheme more accurate results can be obtained, not only for (frustrated) spin models, but also for strongly correlated electron systems. We thus believe that the scheme will play a key role for future state-of-the-art simulations of challenging problems, like the single- and multi-band 2D Hubbard models.

Finally, we point out that there are also other ways to perform a variational optimization with iPEPS than the scheme presented here. A systematic summation of Hamiltonian contributions can also be achieved using the "channel environments" introduced in Ref.~\onlinecite{Vanderstraeten15}. As an alternative to solving a generalized eigenvalue problem at each step, one can also use other approaches to minimize the energy, e.g.~a conjugate gradient method.~\footnote{L. Vanderstraeten et al.,\emph{in preparation}.} Thus, 
there are  more promising options which are interesting to be explored, in order to further improve the performance of iPEPS ground state simulations.


\acknowledgments
The author acknowledges insightful discussions with G.~Vidal on how to use the CTM method to systematically summing up Hamiltonian terms, 
and with L. Vanderstraeten, F. Verstraete, A. Gendiar, and T. Nishino about other variational optimization schemes.  This work is part of the \mbox{D-ITP} consortium, a program of the Netherlands Organization for Scientific Research (NWO) that is funded by the Dutch Ministry of Education, Culture and Science~(OCW). This research was supported in part by Perimeter Institute for Theoretical Physics. Research at Perimeter Institute is supported by the Government of Canada through Industry Canada and by the Province of Ontario through the Ministry of Research and Innovation.


\appendix

\section{Two-site variational optimization}
\label{sec:2site}
In the main part of the paper we discussed how to perform a one-site update, i.e. where one tensor in the unit cell after the other gets iteratively optimized. Similarly as for MPS in one dimension one can also perform a two-site update, where two tensors are updated at once. One advantage of this approach is that the bond dimension can be dynamically adjusted (which is particularly important also when making use of global abelian symmetries~\cite{singh2010,bauer2011}). In the following we describe the main idea how to update, e.g. a horizontal bond involving tensors $A^{[x,y]}$ and  $A^{[x+1,y]}$.

\begin{figure}
\includegraphics[width=0.9\columnwidth]{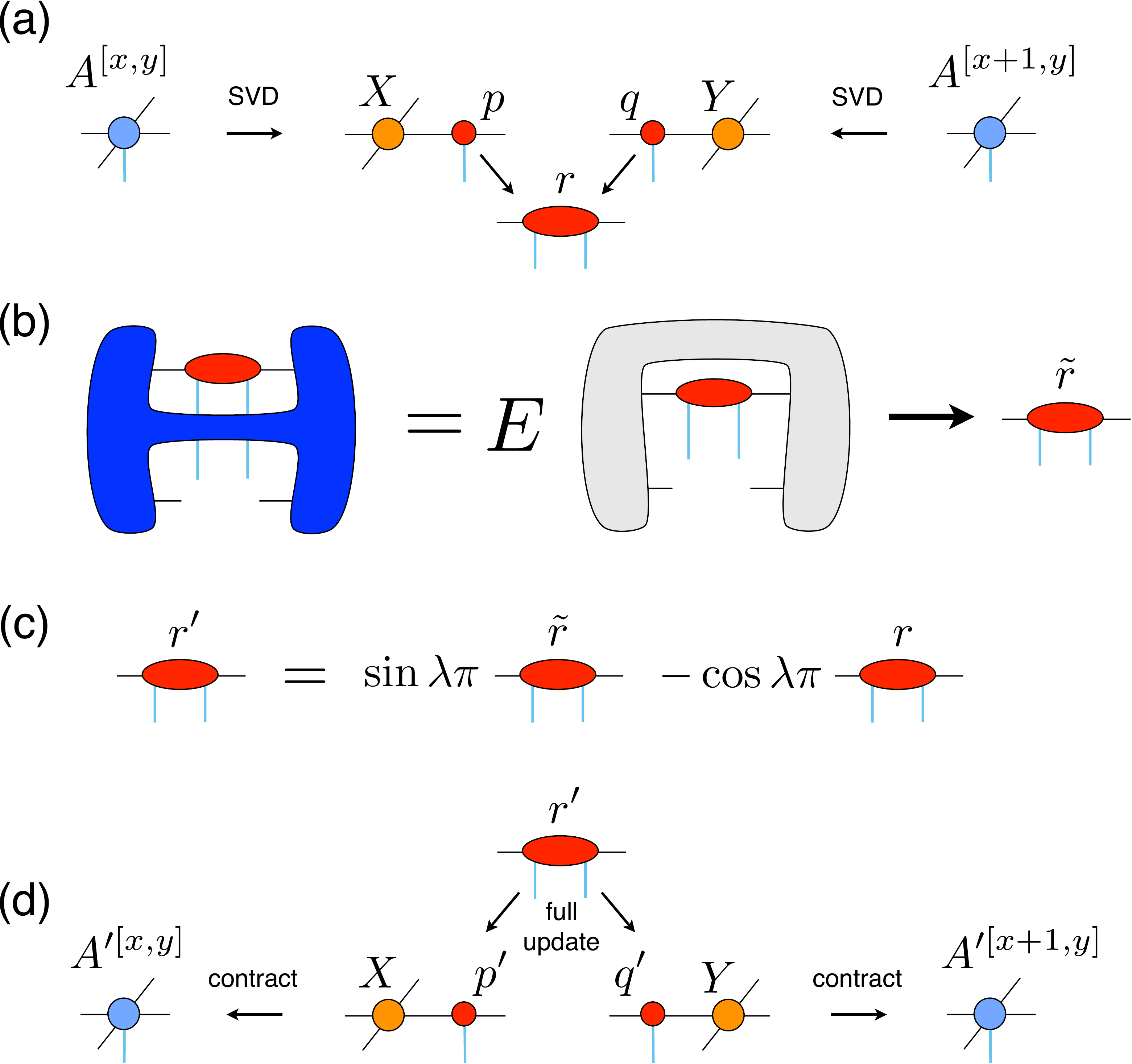}
\caption{(Color online) (a) In order to update the horizontal bond between coordinates $[x,y]$ and $[x+1,y]$, we first split the involved tensors into smaller pieces using an SVD (the singular values are absorbed in the $p$ and $q$ tensors).  (b)~Solving the generalized eigenvalue problem for tensor $r$ yields the solution~$\tilde r$. (c) Ansatz for the solution $r'$ made of a linear combination of $\tilde r$ and the initial tensor $r$, depending on a parameter $\lambda$. (d) The updated tensors $A'^{[x,y]}$ and  $A'^{[x+1,y]}$ are obtained from recombining $p'$ with $X$ and $q'$ with $Y$, respectively, where $p'$ and $q'$ are determined using the full update.
}
\label{fig:twosite}
\end{figure}

Since updating two full tensors at once is computationally expensive, we first split the two tensors $A^{[x,y]}$ and  $A^{[x+1,y]}$ into smaller pieces  using an SVD, shown in Fig.~\ref{fig:twosite}(a). A similar splitting is also used in the simple update and full update to increase the efficiency.~\cite{corboz2010, phien15} The contributions $p$ and $q$ are then combined into a single tensor $r$. We now can formulate the variational optimization with respect to this tensor $r$ by constructing the corresponding ${\bf H}$ and ${\bf N}$ matrices using the H- and norm-environment tensors, similarly as in the one-site update. Solving the generalized eigenvalue problem for the lowest energy eigenvalue yields the tensor $\tilde r$ in Fig.~\ref{fig:twosite}(b). We again make an ansatz for the solution $r'$ by a linear combination of $\tilde r$ and the previous tensor $r$ (Fig.~\ref{fig:twosite}(c)) and determine the best mixing parameter $\lambda$, e.g. with the scheme from Sec.~\ref{sec:practical}. To evaluate the energy one has to decompose $r'$ into two pieces $p'$ and $q'$ (with bond dimension $D$ between them) which can be done using the full update,~\cite{corboz2010} and then use $p'$ and $q'$ to compute the updated tensors $A'^{[x,y]}$ and  $A'^{[x+1,y]}$, respectively, see Fig.~\ref{fig:twosite}(d). 

This completes the update of a horizontal bond, and one can now proceed with updating all the other horizontal and vertical bonds in the unit cell, and perform several sweeps until convergence of the energy is reached.

\bibliographystyle{apsrev4-1}
\bibliography{../bib/refs}

\end{document}